\documentclass[onecolumn]{rsauthor}   
\usepackage{inputenc}
\usepackage[T1]{fontenc}
\usepackage{color}

\usepackage{graphicx}
\usepackage{amsmath}
\usepackage{amssymb}
\usepackage{amsfonts}
\usepackage{amsthm}
\usepackage{endfloat}
\usepackage{endnotes}
\usepackage{setspace}
\usepackage{verbatim}
\usepackage[left=1in, right=1in, bottom=1in, top=1in]{geometry}
\usepackage{times}
\usepackage{helvet}
\usepackage{courier}
\usepackage{bm}
\usepackage{url}
\usepackage{babel}
\usepackage{dcolumn}
\usepackage{multirow}
\usepackage{array}
\usepackage{rotating}
\usepackage{cellspace}
\usepackage{booktabs}
\usepackage[usenames]{xcolor}
\usepackage[numbers]{natbib}

\definecolor{Myorange}{cmyk}{0,0.42,1,0}

\newcommand{\rev}[1]{\textcolor{black}{\textnormal{#1}}} 

\graphicspath{{Fig/}}

\title{Graph analysis of functional brain networks: practical issues in translational neuroscience}

\author{Fabrizio De Vico Fallani$^{1}$, Jonas Richiardi$^{2,3}$, Mario Chavez$^{1}$, Sophie Achard$^{4}$} 

\address{
$^1$Inserm U1127, CNRS UMR7225, Sorbonne Universites, UPMC Univ Paris 06 UMR S1127, Institut du Cerveau et de la Moelle epiniere, ICM, Inria Paris-Rocquencourt, Paris, France. 

$^2$Functional Imaging in Neuropsychiatric Diseases Laboratory, Department of Neurology and Neurological Sciences, Stanford University, USA.

$^3$Laboratory for neuroimaging and cognition, Department of Neurology and Department of Neurosciences, University of Geneva, Switzerland. 

$^4$Laboratoire Grenoble Images-Parole-Signal-Automatique (CNRS-UMR5216), Grenoble, France
}

\corres{Fabrizio DE VICO FALLANI\\
\email{fabrizio.devicofallani@gmail.com}
}

\abstract{
The brain can be regarded as a network: a connected system where nodes, or units, represent different specialized regions and links, or connections, represent communication pathways. From a functional perspective communication is coded by temporal dependence between the activities of different brain areas. In the last decade, the abstract representation of the brain as a graph has allowed to visualize functional brain networks and describe their non-trivial topological properties in a compact and objective way. Nowadays, the use of graph analysis in translational neuroscience has become essential to quantify brain dysfunctions in terms of aberrant reconfiguration of functional brain networks.
Despite its evident impact, graph analysis of functional brain networks is not a simple toolbox that can be blindly applied to brain signals. On the one hand, it requires a know-how of all the methodological steps of the processing pipeline that manipulates the input brain signals and extract the functional network properties. On the other hand, a knowledge of the neural phenomenon under study is required to perform physiological-relevant analysis. The aim of this review is to provide practical indications to make sense of brain network analysis and contrast counterproductive attitudes.
}


\begin{document}

\maketitle

\section{INTRODUCTION}

In the last decade, the use of advanced tools deriving from statistics, signal processing, information theory and statistical physics, has significantly improved our understanding of the brain functioning. Notably, connectivity-based methods have had a prominent role in characterizing normal brain organization \cite{friston_functional_2011} as well as alterations due to various brain disorders \cite{fox_clinical_2010}.

By measuring the magnitude of temporal dependence between regional activities -- using functional modalities such as functional magnetic resonance imaging (fMRI), electroencephalography (EEG), or magnetoencephalography (MEG) -- functional connectivity patterns describe how $N$ different brain areas interact with each other. The resulting $N\times N$ multivariate relationships lead to an interconnected representation of the brain that can be conveniently treated as a network (or graph) \cite{stephan_computational_2000}, leaving space to an holistic approach as a possible alternative to reductionism \cite{mazzocchi_complexity_2008}.

As with other real-world connected systems and relational data, studying the topology of the interactions in the brain has profound implications in the comprehension of complex phenomena, like the emergence of coherent behavior and cognition \cite{varela_brainweb:_2001} or the capability to functionally reorganize after brain lesions (i.e., brain plasticity) \cite{carter_why_2012}. In practice, graph metrics (or indices) such as clustering coefficient, path length and efficiency measures are often used to characterize the ``small-world'' properties of brain networks \cite{bassett.2006.1,Achard2007Efficiencyandcost}. Centrality metrics such as degree, betweenness, closeness, and eigenvector centrality are used to identify the crucial areas within the network. Community structure analysis, which detects the groups of regions more densely connected between themselves than expected by chance, is also essential for understanding brain network organization and topology \cite{Rubinov2010Complexnetworkmeasures}.

The emerging area of complex networks has led to a paradigm shift in the neuroscience community, though many issues remain unaddressed \cite{fornito_graph_2013}. We have only looked at a few aspects of brain networks with rather crude approaches. Most of them rely on single graph metrics which may lack clinical use due to low sensitivity and specificity \cite{stam_organization_2012}, or on mass-univariate link-based comparisons that ignore the inherent topological properties of the networks and yield little power to determine significance \cite{zalesky_network-based_2010}. 
Hence, despite the increasing popularity of graph theoretical approaches to analyze brain connectivity, our understanding of brain organization at the network level is still in its infancy. This is in part due to the fast and wide methodological development of new analytical tools and the inevitably slower rate of absorption by the neuroscience community, which needs to validate their physiological relevance and practical reliability. Another limitation is that researchers developing methods (e.g. mathematicians, physicists, engineers) and neuroscientists (e.g. neurologists, psychologists, psychiatrists) often belong to different scientific domains, with different communities, goals and problems. Although this interdisciplinary integration is fascinating, and eventually leads to fundamental advances in the comprehension of the brain functioning, there is a number of issues that should be considered to optimize collaborative efforts.

Using a common terminology is certainly the first step to improve integration between different scientific fields. This aspect is generally considered as minor with respect to the final goal, whereas more efforts should be made towards an agreed common ``language'' that aims to avoid confusion and make the collaboration efficient. Another important issue is that functional brain networks are the result of a tricky processing pipeline (Fig. 1), which basically consists in improving the quality of the recorded brain signals, constructing the network by means of measures of pairwise functional connectivity, retaining the most relevant links, extracting graph metrics describing topological properties of the network and finally applying statistical procedures to the extracted graph indices to determine differences between populations (e.g. healthy versus diseased) or conditions (e.g. cognitive/motor tasks versus resting states). Each step of this pipeline requires great attention for the selection of the most appropriate method; both theoretical and practical constraints, as well as physiologically relevant hypothesis, should be taken into account. In addition, translating theoretical concepts into practice is not always straightforward. Particularly in the neuroscience, a wide-ranging knowledge and experience across several scientific areas is required to perform proper data analysis and extract informative graph-based neuromarkers that can be used to predict behavior and/or disease \cite{he_graph_2010}. Last, but not least, the neurophysiological interpretation of the obtained results can present difficulties due to the fact that different methods, with increasing level of conceptual abstraction, are applied sequentially to the recorded brain signals and make direct explanation of the underlying neural phenomenon non-trivial (Fig. 2).

In this scenario, we aim to provide a critical review on graph analysis of functional brain networks. Differently from previous reviews in the field, we extend our survey to all the steps of the processing pipeline taking in input the acquired brain signals and giving as output the statistical description of the brain network (re)organization (Fig. 1). Also, we restrict our review on functional brain networks being more susceptible to dynamic reorganization, underlying cognitive/motor tasks or diseases, as compared to structural (or anatomical) brain networks. 
A major thrust of this review is to provide simple and practical indications that can facilitate the use of brain graph analysis and its outcome interpretation by researchers non-expert in the field. We conclude our survey by highlighting the pressing methodological issues, the future challenges, and the impact of technology on the efficacy of graph-based neuromarkers in translational neuroscience.

\section{BRAIN... NETWORKS OR GRAPHS?}
The concept of ``network'' is becoming fundamental in almost all the research fields, from physics to engineering, from social science to neuroscience \cite{BOCCALETTI2006ComplexnetworksStructure}. This has naturally made the word ``network'' refer~\rev{to concepts associated with connected systems that can be very different in their nature, thus generating a non-negligible source of misunderstanding across disciplines. Artificial neural networks, for instance, implement particular machine learning algorithms inspired by Hebbian rules for synaptic plasticity \cite{Bishop1995Neuralnetworkspattern}. The term network is sometimes used in neuroscience to identify distinct brain regions simultaneously active during a generic mental state \cite{Greicius2003Functionalconnectivityin}. 
Here, we refer to networks obtained with brain connectivity estimates, where nodes stand for different brain regions (e.g. parcellated areas or recording sites) and links indicate the presence of an anatomical path between those regions or a functional dependence between their activities.}

In this direction, recent advances in network theory has revolutionized the analysis of brain connectivity patterns estimated from neuroimaging data \cite{bullmore_complex_2009}. However, it should be noticed that complex networks generally refer to real-world connected systems where the links between the nodes are unequivocally defined. Hence, the term network appears justified when considering structural (or anatomical) brain connectivity patterns, where the links represent estimates of real axonal fiber tracks between distant areas, while more caution should be used when considering functional connectivity patterns where links are rather statistical measures of temporal dependence between brain regional activities. In the latter cases, referring to ``graphs'', which do not make explicit assumptions on the nature of the links but rather emphasize the aspect of mathematical modeling, seems to be more appropriate and~\rev{can} be used to improve~\rev{the  interdisciplinary exchange.}
\section{BRAIN NODES}
\label{s:nodes}
The definition of the nodes (or vertices) for brain graphs is modality-specific (Fig. 1 - Nodes). Broadly speaking, voxel-based modalities such as fMRI or PET define nodes in the measurement space (after image reconstruction), while sensor-based modalities like EEG, MEG, or fNIRS offer a choice between assigning nodes directly to sensors or to reconstructed sources.

In voxel-based modalities, \rev{there are several approaches to define brain nodes~\cite{Reus2013parcellationbasedconnectome,Stanley2013Definingnodesin}; here, we only provide a brief overview.} The main choices facing the researcher are the spatial scale most relevant for the analysis (single voxels versus voxel aggregates), and in the case of voxel aggregates the choice of howto combine voxels (parcellation based on prior anatomical knowledge, a data-driven approach, or a hybrid technique~\cite{Pereira2013CreatingGroupLevel}). \rev{Using single voxels directly as brain nodes was the earliest approach for graph analysis~\cite{Dodel2002Functionalconnectivityby,Eguiluz2005ScaleFreeBrain}. Proponents argue that the higher resolution afforded by voxel-level analysis is a better representation of the real underlying system~\cite{Hayasaka2010Comparisoncharacteristicsbetween} and that it allows model-free analysis~\cite{Heuvel2008SmallWorld}, but potential downsides include lower signal-to-noise ratio and increased graph size compared to regional analysis.} Today, \rev{the most commonly used approach is to use a fixed anatomical atlas~\cite{Stanley2013Definingnodesin,Salvador2005NeurophysiologicalArchitecture}}. The type of algorithm used to segment the data into gray matter, white matter, and other tissues (two broad classes are surface-based methods and volume-based methods), typically influences the choice of the atlas, and therefore the ability to compare results across studies.
Unsurprisingly, the number of regions in the atlas, and therefore the number of brain nodes, has a large impact on the extracted graph metrics~\cite{Zalesky2010WholeBrainAnatomicalNetworks, antiqueira_estimating_2010, echtermeyer_integrating_2011}. Thus, topological properties of brain graphs should mostly be compared across similar spatial scales, i.e. similar number of nodes. Additionally, even with the same number of regions, the parcellation resolution in some functionally relevant regions can impact the conclusion of the analysis; for example some atlases have coarser-grained divisions of the occipital lobe, and this could lead to false negatives results when studying the visual system \rev{~\cite{Hammers2003Threedimensionalmaximum,Gousias2008Automaticsegmentationbrain}}.


Another approach \rev{to aggregate voxels into nodes} is to use independent component analysis (ICA)\rev{~\cite{Jafri2008methodFunctionalNetwork}, whereby each independent component is mapped to a brain node, and functional connectivity is computed using component time courses}. Here, no predetermined atlas is needed,
and this type of approach is particularly popular in resting-state studies. Currently some manual intervention
is still needed to sort components into noise and signal components, but steady progress towards automated component sorting is being made~\cite{Perlbarg2007CORSICA,Tohka2008Automaticindependentcomponent,Kundu2013MultiEchoICA}. A further issue is to choose the number of components in the decomposition. Here the field moves mostly by consensus, with around twenty components being used, though recent experiments have used fifty or more~\cite{Kiviniemi2009Functionalsegmentationbrain,Duff2012TaskdrivenICA}. Lastly, a practical issue that
is sometimes overlooked is that ICA algorithms start with a random initialization with no guarantee of reaching a global minimum, and the obtained components can vary across the algorithm's runs. Therefore, \rev{it can be useful} to analyze the stability of components, and to use those components that tend to reappear in different runs of the algorithm. A principled approach to do that is ICASSO~\cite{Himberg2004ValidatingindependentcomponentsICASSO}, whereby the estimated components that tightly cluster together are considered more reliable.

In sensor-based modalities, brain nodes are commonly assigned directly to a sensor or to an electrode \rev{~\cite{Stam2007Graphtheoreticalanalysis,Bassett2009Humanbrainnetworks}}. However, volume conduction in both EEG and MEG causes the signal at each sensor to be a mixture of blurred activity from different inner cortical sources. This effect can either be ignored, in which case brain nodes will suffer from a biased non-neural dependence, or addressed in several ways. This is a well-known problem in the MEG and EEG communities~\cite{Nunez1997EEGcoherency_I}, and there are three broad classes of possible solutions: $i)$ using spatial filters, $ii)$ choosing functional connectivity measures attenuating volume conduction effects, or $iii)$ using cortical source reconstructions. The first class of solutions mainly consists of Laplacian-based techniques removing the signal component shared by groups of neighboring sensors~\cite{bradshaw_spatial_2001}. The second class of solutions consists in using functional connectivity measures taking into account volume conduction effects, such as the imaginary coherence~\cite{Nolte2004Identifyingtruebrain} or the phase lag index~\cite{Stam2007Phaselagindex}. By using the first two approaches, the modeled brain nodes still coincide with EEG/MEG sensors on the scalp, but possible blurring effects are eliminated or, at least, reduced. In the third class of solutions, instead, brain nodes are assigned to sources over a realistic cortex model, through sophisticated reconstruction techniques\rev{~\cite{baillet_electromagnetic_2001,babiloni_multimodal_2003,Schoffelen2009Sourceconnectivityanalysis}
}. However, even after source reconstruction residual blurring effects might still survive, and orthogonalizing signals pairwise is a viable technique that can be further exploited~\cite{Hipp2012Largescalecortical}.
In fNIRS (functional Near InfraRed Spectroscopy), the distance between the emitter and the detector, as well as the head circumference, affect optical path length~\cite{Benaron1995Transcranialopticalpath}. Thus, if brain nodes are assigned directly to sensors, one possible solution is to standardize emitter-detector distances as much as possible, \rev{as was commonly done in early devices~\cite{Scholkmann2014reviewcontinuouswave}}, or to normalize pairwise sensor functional connectivity measures to the emitter-detector distances\rev{~\cite{Tak2014StatisticalanalysisfNIRS}}; furthermore normalizing by subject head size might be necessary to compare subjects with very different head sizes. Finally, brain nodes are typically assigned to sensors or sources, but they can also be assigned to ICA components~\cite{Zhang2010Functionalconnectivityas}.

%
\section{FUNCTIONAL LINKS}
\label{s:links}
After defining the brain nodes, assigning the links between them is the subsequent crucial modeling step (Fig. 1 - Links). In functional neuroimaging the links of a brain graph are given by evaluating the similarity between two brain signals, through functional connectivity (FC) measures~\cite{fingelkurts_functional_2005}. These data-driven methods do not require modeling assumptions and can be generally applied to brain signals coming from all the neuroimaging techniques (e.g. fMRI voxels, EEG/MEG sensors, cortical sources). Alternative model-based methods (often referred as effective connectivity) can be used when realistic hypotheses of putative connectivity schemes are available from previous neurophysiological evidence ~\cite{friston_functional_2011}. In these cases, the number of distinct brain nodes that can be involved in the model is however rather small, leading to relatively simple connectivity patterns that can be visually described without the use of graph theoretical approaches.
In the last decade many methods have been proposed to measure FC based on different principles ranging from signal processing to autoregressive modeling and information theory. Each method has its own specifications but essentially they all apply to a set of signals $X_1(t) \ldots X_N(t)$ recorded from $N$ different brain nodes, where $N \geq 2$,~\rev{and give, for each pair of nodes $i$ and $j$, the magnitude of the pairwise statistical relationship between their activities. These magnitudes represent the weighted links $w_{ij}$ of the brain graph.
}

FC methods fall into two broad categories: those measuring symmetric mutual interaction (undirected weighted links) and those measuring asymmetric information propagation (directed weighted links). Within these two categories, further classifications can be made according to their ability to measure linear or nonlinear relationships, to consider bivariate or multivariate interactions, and to apply in the frequency domain. The specific technical requirements, like the need for signal stationarity or the selection of appropriate parameters, also represent distinctive traits. Table~\ref{t:connMeasures} presents a list of the basic theoretical principles that are implemented in the large part of the existing FC methods.~\rev{Most of them give a positive real value that weights the interaction $w_{ij}\in\mathbb{R}^{+}$ between the brain node $i$ and $j$. The higher the weight, the stronger the FC between the brain nodes is. Other methods, based on the concept of correlation, can give negative values reflecting anti-correlated dependencies. In this case, a possible procedure in brain graph analysis is to consider the absolute value of the resulting correlation coefficient~\cite{Achard2006JNeuro}, assuming that what is relevant is weighting the presence of a statistical interaction, regardless its sign, which could also present difficulties in terms of neurophysiological interpretation~\cite{meunier_age-related_2009,buckner_human_2010}, i.e. one cannot directly associate inhibitory/excitatory interaction between brain signals on the basis of the sign of the estimated correlation coefficient~\cite{liang_anticorrelated_2012}. 
}
The term functional connectivity is used as a generic, often ambiguous, description of the link weight between two brain nodes. It is often claimed that two brain regions ``have a high functional connectivity'' or ``are strongly functionally connected''. However, it is fundamentally different stating, for example, that two brain nodes have a high phase coherence or Granger-causality. A Granger-causality does not necessarily imply a phase coherence and these two methods can give completely different link weights even when applied on the same brain signals~\cite{schelter_handbook_2006}. It seems therefore recommendable referring to the exact theoretical principle implemented by the FC method in order to avoid generalization and facilitate outcome interpretation.

Nowadays, several FC methods are available and, likely, many others will be developed in the future. Differently from  brain nodes, whose definition is mainly determined by the used neuroimaging technique, the definition of the links is more tricky due to the large number of available FC methods that can be used. Thus, the choice of the most appropriate FC method is not as trivial as it would seem at first glance.
By making a parallel with social networks, where nodes stand for different people, it would be equivalent to wonder if it's better to weight the links as a function of the number of exchanged emails or of the times they have physically meet each other. Both these methods aims to measure the level of acquaintance of people, yet people can exchange a lot of emails (i.e. high connectivity) while they have never meet each other (i.e. null connectivity). 

A few studies that have attempted to compare a subset of FC methods with respect to their ability to correctly detect the presence of simulated connectivity schemes in a multivariate data set ~\cite{quian_quiroga_performance_2002,david_evaluation_2004,ansari-asl_quantitative_2006}. Unfortunately the results are not unequivocal and the performance of the measures seems to depend upon characteristics of the data set as well as characteristics of the method itself. A possible approach may be then to use several methods and search for measures that are as consistent as possible~\cite{stam_organization_2012}. However, this procedure seems highly time-consuming and, above all, lacking of a precise rationale. A more reasoned approach would consist in predetermining the FC method according to plausible hypothesis related to the experimental study.
When the scientific goal is clear and the experimental protocol is well designed, the choice of the FC method can be often a natural consequence that can be selected \emph{a priori}. To make an example, let us consider the case of epileptic seizures, which are episodes of abnormal excessive neuronal activity leading to convulsion and/or mild loss of awareness. In temporal lobe epilepsy, most of the seizures begin as focal and rapidly become generalized (diffused) for several seconds. If we were interested in identifying the epileptic foci during a generalized epileptic seizure, the use of an undirected FC method (e.g. correlation-based) would not be a good choice as the large part of the brain signals will be highly correlated and all the nodes in the graph will be strongly connected, thus making difficult to retrieve possible propagation schemes. Instead, a directed FC method (e.g. causality-based) would be theoretically more informative due to its ability to measure activity propagation and then identify sources of information flows~\cite{wilke_graph_2011}. Nevertheless, if our goal aimed to measure synchronization between brain regions during interictal periods, then an undirected FC method (e.g. based on phase coherence), would be appropriate for our brain graph analysis~\cite{chavez_functional_2010}.

Further elements playing a role in the selection of the FC method depend on the neuroimaging technique that is used to record brain activity. Basically, we can distinguish between techniques capturing physical phenomena associated with neural processes that have either high temporal dynamics (e.g. EEG/MEG in the order of milliseconds) or low temporal dynamics (e.g. fMRI, in the order of seconds). Both scalp EEG and MEG signals exhibit distributed physiological changes during motor/cognitive tasks coded in specific frequency bands ranging from slow (Theta $\lesssim 7$~Hz) to rapid oscillations (Gamma $\gtrsim 30$~Hz). These changes can involve either local and distant brain regions and include increase of Alpha ($8-13$~Hz) activity during mental calculation~\cite{palva_new_2007} or decrease of Beta ($14-20$~Hz) activity during the performance of motor acts~\cite{mcfarland_mu_2000}. Higher frequency changes ($\gtrsim 80$~Hz) can even be observed with intracortical EEG recordings during epileptic seizures that reflect the activity of very small neuronal generators~\cite{jacobs_curing_2009}. In the case of EEG/MEG signals the use of time-domain FC measures will give brain graphs that cannot inform on the interactions at different frequency bands. Frequency-domain methods would be instead more suitable due to their ability to extract brain graphs at multiple frequencies. Time-domain FC methods seem instead more appropriate for fMRI signals, which present physiological as well as pathological changes in a lower and restricted frequency range ($\leqslant0.1$~Hz)~\cite{fox_spontaneous_2007}. In this case, the use of wavelet-based frequency filters is also particularly suitable for isolating and investigating self-similar (fractal) properties of functional connectivity \cite{Bullmore2004Wavelets,achard_fractal_2008}. 

Brain graphs can also be modeled from electrophysiological recordings of cortical \textit{in-vivo} or \textit{in-vitro} neuronal ensembles. At this microscopic spatial scale, nodes are associated with single neurons, and functional links are measured by the statistical analysis the cells' action potentials~\cite{Humphries_spikeCommunities2011}. As neither the shape of the action potential nor the background activity seems to carry relevant information, neuronal responses are normally reduced to binary signals (spike/no spike) where the only maintained information is the timing of the event. Although the extension of FC methods to binary signals is not straightforward, some measures of spike synchrony (or spike train distances) have been developed for multivariate datasets~\cite{kreuz_spikeSynchrony2007, kreuz_spikeSynchrony2009}. Because of the binary nature of these signals, such methods can be mainly considered as statistical tests for neuronal codes (ranging from rate codes to coincidence detectors) and are therefore time-scale dependent.

FC methods rely on statistical definitions and therefore require specific assumptions on the nature of the signals to obtain reliable measures. In general, the information that can be extracted from a noisy signal depends on the number of its time points, i.e. the higher the number of time points, the more reliable the information extracted is. Intuitively, when two or more signals are considered the number of available time points should be even higher to take into account possible interaction effects. This suggests that FC measures present important bias which depends on the length of the recorded signals. From a neurophysiological point of view, dealing with sufficiently long brain signals ensures that the neural process under investigation can occur several times and that it can be sufficiently represented. This requirement can be easily obtained during no-task or resting state (RS) experimental protocols. RS conditions elicit a characteristic spontaneous brain activity which is distributed according to a complex network (i.e. default-mode network), whose topology reveals fundamental information about the brain organization and informs on a variety of mental and developmental disorders, including schizophrenia and autism~\cite{buckner_brains_2008}. When the study consists in extracting information from oscillatory activity at a given frequency, the length of the brain signals should be adjusted to the targeted frequency range \cite{Lachaux1999}. In~\cite{Birn2013Timeseries} the authors advocate to use at least $12$ minutes of fMRI signal acquisition with a repetition time of $2.6$~s and a $0.01-0.1$~Hz band-pass filter for the functional connectivity between eighteen brain regions. When considering task-based (TB) conditions, brain signals can exhibit relatively rapid changes underlying motor or cognitive behaviors and short time windows are generally applied to focus on transient phenomenona. In these cases, the number of available time points can diminish significantly and multiple repetitions of the experiment (i.e. trials or epochs) are commonly performed to increase the total number of the data points. The brain graph can then be obtained by integrating the information across trials with the same length \cite{WibralMEG}. The simplest way to do that is computing the FC after concatenating the trials or averaging the FC measures obtained from each single trial; more sophisticated approaches consist in exploiting the statistical variability of the trials to improve the final FC measure~\cite{schelter_handbook_2006}. 

The above indications can be considered valid for linear and bivariate FC measures. When there is valid hypothesis on the existence of possible non-linear interactions (e.g. epileptic crisis, movement onset), non-linear FC methods should be used and a higher number of data points is needed to improve the characterization of rapid changes~\cite{netoff_detecting_2006}. A higher number of data points is also required when dealing with multivariate FC measures~\cite{pereda_nonlinear_2005, AmblardMichel2010}. Multivariate methods have been introduced to avoid the emergence of spurious FC (i.e. false positives) between two signals that is due to the presence of other signals interacting with them. These methods are based on the computation of joint conditional probabilities (e.g. partial correlation~\cite{marrelec_partial_2006}, mutual information~\cite{chai_exploring_2009}) or the estimation of a large number of parameters (e.g. DTF~\cite{kaminski_new_1991}, PDC~\cite{baccala_partial_2001}) from the entire set of signals. \rev{Unbiased methods have been also developed for multivariate phase coupling measures~\cite{vinck_improved_2011}
}. 

Finally, almost all the existing FC methods require the stationarity or quasi-stationarity of the recorded brain signals~\cite{kwiatkowski_testing_1992}. Stationarity implies that signal properties, like the mean and the variance, do not change in time. The stationarity property is generally satisfied in healthy RS conditions, where no particular changes in the mean and variance of the signals are expected. However, possible departure from stationarity can occur in TB or diseased RS conditions~\cite{fingelkurts_functional_2005}. In those cases it's a duty of the researcher to "reestablish" the stationarity of the signals by either performing appropriate pre-processing (e.g. detrending) or considering shorter time windows where the mean and variance of the signals remain stable. Other solutions consists in using wavelet transformations, which are able to make stationary very complex signals such as fractal or long memory signals~\cite{Bullmore2004Wavelets} or applying time-varying FC methods which do not require stationarity (e.g. time-frequency measures or adaptive multivariate methods)~\cite{sanei_fundamentals_2013}.
Regressing the global mean out of the time-series~\cite{Macey2004methodremovalglobal} is a more controversial approach~\cite{Murphy2009impactglobalsignal,Chai2012Anticorrelationsinresting,Saad2012TroubleatRest} as it can introduce negative correlations~\cite{Murphy2009impactglobalsignal,Weissenbacher2009Correlationsandanticorrelations}. In low-noise conditions the global variation of the blood-oxygen-level dependent (BOLD) signal may have significant correlation with the brain signals of interest, such as those constituting the fMRI default-mode network~\cite{Chen2012methodtodetermine}. On the other hand, global signal regression has also been shown to increase specificity of the link weights~\cite{Weissenbacher2009Correlationsandanticorrelations} and can help mitigate the effects of head motion~\cite{Power2014Methodstodetect}.\rev{Indeed, head motion is a major cause of non-stationarity with a non-negligible impact on the brain graph topology. Two broad classes of approaches emerge here: regression and deletion. For regression, an early and effective approach used originally in TB studies~\cite{Friston1996MovementRelatedEffects}, and now commonly used in RS conditions, includes realignment parameters of the fMRI volumetric signals as nuisance covariates in a linear model, and uses prediction residuals as estimates of the denoised signal. A recent update on this technique consists in combining the intensity spike-specific regressors with the derivatives and squares of the realignment parameters~\cite{Satterthwaite2013improvedframeworkconfound}. Another approach is to add a de-spiking step prior to motion regression~\cite{Patel2014waveletmethodmodeling}. For deletion, it has been proposed to drop time points that have excessive motion~\cite{Power2012Spuriousbutsystematic}. This has the disadvantage of potentially yielding different signal length between subjects, and irregular sampling.
}

Many toolboxes are publicly available to compute FC from brain signals. Here we refer to those developed under the Matlab\textsuperscript{\textregistered} environment and the free R software. For multivariate analysis of FC, the BSMART\footnote{\url{http://www.brain-smart.org/}} toolbox provides a plethora of estimators such as partial power, coherence, partial coherence, multiple coherence and Granger causality. Similarly, eConnectome\footnote{\url{http://econnectome.umn.edu/}} is an open-source MATLAB software package for estimating brain FC from electrophysiological signals.  A toolbox for Granger causality analysis (both in time and frequency domain) is also provided by Anil K. Seth\footnote{\url{http://www.sussex.ac.uk/Users/anils/aks_code.htm}}. An integrated toolbox devoted to the advanced processing of EEG and MEG data is FieldTrip\footnote{\url{http://fieldtrip.fcdonders.nl/}}. It includes many  graphical user interfaces for data preprocessing, source estimation, FC analysis and data visualization. The SPIKY\footnote{\url{http://wwwold.fi.isc.cnr.it/users/thomas.kreuz/sourcecode.html}} toolbox also provides a graphical user interface with several estimators for spike synchrony or spike distances. An R-based toolbox to compute wavelet correlation is available in the Brainwaver\footnote{\url{http://cran.r-project.org/web/packages/brainwaver/index.html}} package, also available for MATLAB environment in the Connectivity Decoding Toolkit\footnote{\url{https://www.stanford.edu/~richiard/software.html}}.

\section{GRAPH FILTERING}
\label{s:threshold}
 \rev{Computing FC between all the pairs of brain nodes leads to $N\times N$ relationships that can be conveniently represented in a full weighted matrix $\textbf{W}_{N\times N}$, containing all the pairwise FC measures $w_{ij}$ corresponding to the weighted links of the brain graph. When using directed FC methods (see Section 4) the resulting $\textbf{W}_{N\times N}$ is asymmetric with a total number of weighted links equal to $N(N-1)$, excluding self-connectivity. The weighted matrix is instead symmetric when using undirected FC methods and the total number of links is $N(N-1)/2$.
} 
\rev{Some authors have proposed to consider all the available information from $\textbf{W}_{N\times N}$ and keep all the available weighted links~\cite{Ginestet2011Brainnetworkanalysis,Rubinov2011NegativeWeight}. 
}
However, as described in the previous section, FC measures can be affected by various non-neural phenomena, including spatial parcellation (used in fMRI or source reconstruction studies), noise characteristics, and peculiar signal properties like, for example, the long memory or fractal nature of time series~\cite{achard_fractal_2008}. Thus, possible difficulties related to the statistical uncertainty on the link weights can arise when interpreting the resulting extracted graph metrics. 
\rev{An alternative approach consists in filtering the matrix $\textbf{W}_{N\times N}$ and maintaining only those links whose weight corresponds to FC measures different from zero. A common way to do that, is to fix a threshold $T\in\mathbb{R}^{+}$ and remove the links whose weight $w_{ij}$ is lower than $T$ (practically, this is done by setting the corresponding entries in $\textbf{W}_{N\times N}$) to zero. The weights of the survived links can be either maintained or transformed into binary values by setting them to one in $\textbf{W}_{N\times N}$~\cite{Rubinov2010Complexnetworkmeasures}.
Using graph filtering should be used cautiously, as it allows only to compare the topological organization of the strongest connected pairs of brain nodes~\cite{Fornito2012Neuroimage}. 
} 
There has been a long debate on how to fix the threshold and we could distinguish between two possible strategies related to statistical and graph theoretical (or topological) arguments (Fig. 1 - Threshold).

From a statistical point of view, some properties of the FC measures (e.g. the shape of their distribution under the null hypothesis) can be either derived theoretically, from the definition of the FC method itself, or generated by using data surrogates. This information is exploited to fix a statistical threshold above which the percentile corresponding to the FC measure can be considered significant, i.e. distant from the null hypothesis of no-connectivity.
For example, FC measures based on correlations or partial correlation can be assessed using asymptotic statistics~\cite{Muirhead1982MultiStat,Whitcher2000WaveCovariance}. 
Other statistical tests, currently used in phase coherence analysis (e.g. the Rayleigh statistics), are particularly robust when the number of available trials is larger than 50~\cite{fisher_Book89}. The use of data surrogates is also a good data-driven solution for the statistical assessment of FC measures in presence of short signals~\cite{schelter_handbook_2006}.
\rev{Other approaches allow to determine the probability that correlation-based FC measures are significantly higher than what expected from independent signals by using the Fisher's Z transformations~\cite{Jenkins_Book1968,valencia_fMRI_communities2009}. 
Positive biases have been also reported for common FC nonlinear measures, such as phase synchrony~\cite{chavez_towards_2006,aydore_note_2013}
}. 
Another important issue comes from the fact that statistical significance should be tested for each link. In a directed brain graph with $N$ nodes there are $N(N-1)$ links and then $N(N-1)$ tests to perform. In statistics, this issue is called multiple testing and possible solutions consist in adjusting the statistical threshold to number of tests. In neuroimaging, if the number of the brain nodes is high and the length of the signals is short, high correlations may emerge simply by chance~\cite{Hero2011Correlation}. This suggests that the threshold should be chosen according to the dimension of the problem, i.e. number of brain nodes and length of the signals. 
For example, it can be demonstrated that for a brain graph with 500 nodes and a length of the fMRI signals equal to 500 time points, the measured correlations greater than $0.567$ are statistically significant at the third wavelet scale~\cite{Achard2006JNeuro}). For the sake of completeness, we report in table~\ref{t:filterGraph} the typical correlation weights that are significantly different from zero when applying family wise error rates (FWER) and pre-filtering wavelet approaches that keep the third scale. 
Recently, alternative approaches have been proposed that reduce the number of statistical tests and mitigate the multiple testing issue. In particular, the network-based approach  consists in grouping the brain nodes into subnetworks and perform the tests at the subnetwork level~\cite{Zalesky2010NBS}. The subnetwork-based analysis relies on the  knowledge of the brain anatomical structure. It can be demonstrated that such \textit{a-priori} information leads to an increase in power for the multiple testing of the significant brain links~\cite{Meskaldji2011Statconnectome}.  

\rev{From a topological perspective, the number of existing links influences most of the topological metrics that can be extracted from a network~\cite{Kaiser2011Neuroimage}. For example the shortest path length is monotonically decreasing when adding links to a connected network. As a consequence, possible differences between group of subjects/conditions, could emerge because a different connection density. A possible solution is then to fix a threshold $T$ on the number of links among the ones that are tested as significantly different from zero, rather than on the weights~\cite{Kaiser2011Neuroimage, Fornito2012Neuroimage}. In this case, the group/condition analysis can be performed by comparing brain graphs with the same number of links and possible emerging differences can be associated with a different network topology, excluding any network density effect.}
However, choosing the density threshold of the brain graph may not be sufficient to reveal robust topological changes between subjects/conditions. It has already been observed that when the density threshold is too high or too low, the brain graph cannot be differentiated from a random or lattice network~\cite{Achard2007Efficiencyandcost}. Then, a common procedure is to choose a sequence of increasing threshold within a range of admissible densities with respect to random or lattice networks. The comparison between groups/conditions is then performed exploiting the entire collection of predetermined thresholds. A first approach consists in testing the difference between the topological properties of the brain graphs at each threshold \cite{Bassett2012Alteredrestingstate}; a second approach is to compare the properties of the brain graphs by integrating the topological metric over a collection of available thresholds, as in \cite{Ginestet2011Brainnetworkanalysis}; a third approach consists in comparing the profile of the properties of the brain graphs across the threshold range, measuring, for example, the area under the curve \cite{Bassett2012Alteredrestingstate}.

\section{TOPOLOGICAL METRICS}
\label{s:index}
Topological properties of brain networks can be derived at the large-scale of the whole brain, i.e. metrics on the entire graph; at the intermediate-scale of several regions of the brain, i.e. metrics on subgraphs; or locally, at the small-scale of single regions, i.e. metrics on single nodes (Fig. 1 - Metrics). A vast number of graph metrics do exist to extract topological properties~\cite{Rubinov2010Complexnetworkmeasures, Sporns2011NetworksBrain} and, depending on their definition, they can apply to brain graphs with different features and/or require specific assumptions (see Table~\ref{t:graphMetrics}). \rev{Notably, the definition of some graph indices may carry mechanistic biases and particular attention should be paid in the way they are used or interpreted. The well-known signature of hierarchical network structure, for instance, has been shown to be a consequence of degree-correlation biases in the clustering coefficient definition~\cite{soffer_network_2005}; the identification of high-degree nodes (or hubs) in brain graphs obtained with correlation-based FC measures could be explained by the size of the subnetwork they belong to~\cite{power_evidence_2013}.
It follows that the selection of the topological metrics to extract from the brain graph is an important step that requires great attention. Depending on the nature of the neuroimaging experiment, the FC method, and the filtering threshold, some graph indices can result more appropriate than other ones. For example, the characteristic path length, which evaluates the minimum distance between nodes, is not robust for networks with disconnected nodes and the global-efficiency should be used instead~\cite{Latora2001EfficientBehaviorSmall}.
} 

Interesting clinical insights into several brain diseases have been obtained using graph analysis of functional brain networks~\cite{Guye2010GraphtheoreticalAnalysis,stam_organization_2012}. At the large topological scale, the small-world organization, whereby both integration (relatively high global-efficiency/low path length) and segregation (relatively high local-efficiency/clustering coefficient) of information between brain regions are supported, is significantly altered by several pathologies such as schizophrenia~\cite{Liu2008Disruptedsmallworld, Wang2010Impairedefficiencyfunctional}, autism~\cite{barttfeld_big-world_2011}, stroke~\cite{Wang2010Dynamicfunctionalreorganization}, spinal cord injuries~\citep{fallani_cortical_2007} and Alzheimer's
disease~\cite{Supekar2008Networkanalysisintrinsic}. 
Looking at intermediate topological scales complementary information can be obtained by focusing the analysis on subgroups of brain nodes. \rev{In~\citep{chavez_functional_2010}, a network clustering approach has revealed that brain graphs of epileptic patients suffering from absence seizures have a more regular modular organization compared to healthy subjects. In stroke, where typically one hemisphere is affected, forcing the analysis on two specific subgroups coincident with the two hemispheres, has allowed to determine that the observed decrease of global small-worldness is mainly caused by an abnormal increase of interhemispheric connectivity and that the connection density of the affected hemisphere correlates with the residual motor ability of patients~\cite{DeVicoFallani2013Multiscaletopologicalproperties}.}
At small topological scales, the centrality of a brain node (its propensity to act as a ``hub'') can be measured in several ways. For example, node
betweenness measures the number of shortest paths passing through the node. This metric is of particular interest to study brain syndromes where disconnection effects are expected. For example, it has been shown that central nodes in healthy subjects are good predictors of atrophy for several neurodegenerative diseases~\cite{Zhou2012Predictingregionalneurodegeneration}, and that subregions of the cingulate cortex \rev{(belonging to the default-mode network)} are less connected to \rev{other} resting-state networks (lower participation coefficient) for more severely demented patients~\cite{Brier2014FunctionalconnectivityGraphTheory}. \rev{A recent study has also pointed out the importance of nodal centrality metrics, over larger-scale metrics, in characterizing the brain graph reorganization in comatose patients and proving network-based neuromarkers that can be used to evaluate consciousness states~\citep{Achard2012Hubsbrainfunctional}.}

Selecting the appropriate topological metrics to compare brain graphs depends, primarily, on the research question and, secondly, on practical computational requirements. If the question of interest is at the level of the whole brain, then large-scale graph indices should be used. However, this will yield information about neither subgroups nor single brain nodes or links. If a finer-grained topological description of the brain graph is desired, intermediate- or small-scale graph metrics should be used (see Table~\ref{t:graphMetrics}). Thus, given prior knowledge, either a specific set of graph indices reflecting the hypothesis of interest can be computed, or a larger number of topological metrics can be computed and input to (un)supervised learning algorithms to extract the most relevant ones. 
\rev{As a side note, we notice that for large brain graphs the computation of some topological metrics (e.g. network motifs~\cite{Milo2002Networkmotifssimple}, weighted local-efficiency~\cite{latora_economic_2003}) could become highly intensive and, sometimes, impractical.
}
While standard graph theory has mainly focused on unweighted (or binary) networks (e.g. social networks), the extension of graph metrics originally developed for those binary networks to the weighted case is not always straightforward. Particular attention should be paid for brain graphs with negative link weights (such as those obtained from correlation-based FC measures) as very few topological metrics have been defined for their characterization~\cite{Rubinov2010Complexnetworkmeasures}. Moreover, since brain signals recorded from neuroimaging techniques are typically noisy, link weights could be affected by non-neural contributions (see Section~\ref{s:links}). In this case, two brain graphs with an identical topology (i.e. same distribution of links), but with different noisy weights, could appear different when examined through the prism of weighted graph metrics. Finally, it is worthwhile to mention that, under specific conditions, weighted graph indices can degenerate towards simple summary metrics of the network. For example, if the dynamic range of the link weights is small, the weighted global-efficiency is equal to total strength (i.e. the sum of all the weighted links) of the network~\cite{Ginestet2011Brainnetworkanalysis}. A common solution to avoid the above issues, is to average the extracted graph indices across a range of thresholds~\cite{Ginestet2011Brainnetworkanalysis} or, more simply, neglect the weights and use topological metrics defined for unweighted networks. 

Moving to weighted brain graphs also implicates more conceptual issues. As stated in Section~\ref{s:links}, the link weights of a brain graph give a measure of similarity between the signals of two brain nodes.~\rev{This similarity can be conceptually regarded as a sort of functional proximity, but this conceptualization contrasts with the original definition of weight in real networks that is typically related to physical distances, e.g. highway networks. Generally, the concept of distance in brain graphs is controversial because of its actual interpretation. While it is intuitive to interpret a weighted link between two brain nodes as a significant interaction between their signals, it appears a little less intuitive to extend this interpretation when the same two brain nodes are connected through a longer sequence of differently weighted links involving other nodes.} Hence, a possible solution to compute distance-based graph metrics (e.g. the characteristic path length or the global-efficiency) from weighted brain graphs, is to substitute the link weights with their reciprocal $1/w_{ij}$~\cite{fallani_cortical_2008,Rubinov2010Complexnetworkmeasures}. Other possibilities consist in applying weight-to-distance transformations ensuring triangular inequality. For example, in the case of correlation-based FC measures, the nonlinear function $\sqrt{2(1-w_{ij})}$ can be applied to transform the link weights of the brain graph into distance metrics~\cite{mantegna_hierarchical_1999}. Eventually, the use of thresholding procedures (see Section~\ref{s:threshold}) for filtering the original brain graph into a sparse binary graph can represent valid alternatives to avoid misuses of distance-based topological metrics. 

\rev{Many toolboxes are publicly available to compute graph indices from brain connectivity matrices. The Brain Connectivity Toolbox (BCT)\footnote{\url{http://www.brain-connectivity-toolbox.net/}} contains a large number of topological metrics implemented in Matlab\textsuperscript{\textregistered}. It is largely used in the neuroscience community and most of those graph indices are also available in C++. 
Other freely available libraries include: $i)$ MatlabBGL\footnote{\url{http://dgleich.github.io/matlab-bgl/}}, the most complete and robust Matlab package for working with relatively large sparse networks (with hundreds of thousands of nodes); $ii)$ Matlab Tool for Network Analysis package\footnote{\url{http://strategic.mit.edu/downloads.php?page=matlab_networks}}, available at the  Massachusetts Institute of Technology it includes a collection of graph indices and several routines for graph visualization; and $iii)$  iGraph\footnote{\url{http://igraph.sourceforge.net/}}, a toolbox freely available in R, Python and C that implements all the essential topological metrics used in brain connectivity analysis. 
}
\section{STATISTICAL ANALYSIS}
\label{s:stats}
Comparing brain graphs in terms of topological metrics is mostly done 
 $i)$ by comparison with theoretical reference graphs generated by simulation models, or $ii)$ by comparing set of brain graphs, either between different experimental conditions (e.g. task versus rest) or different populations (e.g. diseased versus healthy).

Comparison with reference graphs has been used to normalize the topological metrics computed on brain graphs with respect to a ``null model''. Reference graphs have the same number of nodes and links of the brain graphs but they have a different topology, which is typically random or regular~\cite{Watts1998Collectivedynamicssmall}. Random configurations are generally preferred because they have known statistical properties and can ideally simulate the absence of a neural organizational principle ~\cite{bullmore_brain_2011}.
A critical issue in this approach is how to select the proper rewiring algorithm, which typically can preserve, or not, some properties of the original brain graph~\cite{Rubinov2010Complexnetworkmeasures}). 
In general it is often desirable to preserve the original degree distribution as, for example, random graphs generated for a brain graph exhibiting a scale-free organization~\cite{Eguiluz2005ScaleFreeBrain} will maintain such characteristic property. However, details of the rewiring algorithm matter in order to uniformly sample from the space of random graphs~\cite{Artzy-Randrup2005Generatinguniformlydistributed}. An early result on MEG data has used the Watts-Strogatz random rewiring algorithm to generate equivalent reference graphs and reported evidence of small-worldness in low and high frequency bands by computing clustering coefficient and characteristic path length~\cite{Stam2004Functionalconnectivitypatterns}. However, recent results on anatomical brain networks show that small-world properties can change dramatically depending on the choice of null model~\cite{Hosseini2013Influencechoicenull}.
In addition, selecting reference graphs representative of appropriate null hypothesis can also depend 
on other design choices, like the FC method used to construct the brain graph. For example using correlation-based FC measures can lead to artificially inflated clustering coefficients when compared to standard random graphs and the use of the Hirschberger-Qi-Steuer null model, keeping the distributional properties of the brain covariance matrix, has been advocated~\cite{Zalesky2012usecorrelationas}. More generally, it can be argued that false small-worldness can also emerge when using bivariate FC methods (see Section~\ref{s:links}), as they do not distinguish between direct and indirect signal influences and can lead to spurious links~\cite{kus_determination_2004}, thus favoring clustering increases. In these cases, the use of statistical thresholds, rather than density or weight thresholds (see Section~\ref{s:threshold}), has proved to reduce this artificial occurrence, at least in Granger causality-based FC methods~\cite{toppi_how_2012}.

Between-condition statistics on brain graph metrics have provided empirical evidence for psychological models. One recent example using MEG, in support of the global workspace theory, shows that local-efficiency decreases with the increment of the cognitive load, leading to lower local clustering and a more segregated brain graph~\cite{Kitzbichler2011Cognitiveeffortdrives}. Using similar representation and inference approaches, between-group statistics on brain graph indices has been very successful at showing differences between different populations. For instance, global and local-efficiency of fMRI-derived graphs in elderly subjects has been found to be significantly lower than in young subjects~\cite{Achard2007Efficiencyandcost}.
Analyzing clinical populations is particularly illuminating, in particular for diseases that can be 
described as disconnection syndromes~\cite{Geschwind1965Disconnexionsyndromes}. Schizophrenia leads to a significant increase of the path length, compared to healthy subjects, that is significantly correlated with disease duration~\cite{Liu2008Disruptedsmallworld}. This supports the idea that integration between brain regions is deficient in schizophrenia. 

Despite their popularity, computing topological metrics to compare brain graphs across conditions/populations requires a certain number of assumptions and precautions. In particular, imposing a fixed number of brain nodes, as well as their ordering, will avoid the computationally demanding node correspondence
problem. As mentioned in Section~\ref{s:nodes}, this can be achieved for fMRI and source imaging studies by using a common parcellation for all subjects and conditions, while it is not necessary for sensor-based modalities. Also, many graph metrics depend (non-linearly) on the number of links in the brain graph, while the total strength of the network can yield different weighted graph properties, with no changes in the underlying topology. For all these cases, we refer to the possible solutions described in Section~\ref{s:threshold} and~\ref{s:index} .

Assuming properly measured brain graph indices, three related approaches can be used to statistically manipulate topological metrics of different conditions/populations (Fig. 1 - Stats): $i)$ hypothesis testing, where the emphasis is on the group differences, $i)$ statistical modeling, which focuses on determining possible relation with behavioral outcome, and $iii)$ classification, stressing the use of machine learning approaches to separate groups of brain graphs. 

Hypothesis testing (e.g. T-test), and more in general statistical models (e.g. general linear models), require the Gaussianity of the input data. However the distribution of several graph metrics is non-Gaussian and may also depend on the spatial resolution of the brain nodes (in fMRI: regions of interest versus voxels)\rev{~\cite{Fornito2010Networkscalingeffects,Hayasaka2010Comparisoncharacteristicsbetween}}. For example, the global-efficiency is always bounded between $0$ and $1$, while the degree distribution of brain graphs with scale-free configurations are described by a (exponentially truncated) power law~\cite{Joyce2010newmeasurecentrality}. In both cases, there is a clear departure from Gaussianity.~\rev{Possible solutions to circumvent this situation consist in normalizing the brain graph indices by those obtained from equivalent reference graphs by using Z-score transformations~\cite{fallani_cortical_2008,alexander-bloch_anatomical_2013}. Otherwise, more flexible techniques, including non-parametric tests (e.g. Wilcoxon test, Kruskall-Wallis) and generalized linear models, should be used despite the potential loss of power and precision~\cite{zar_biostatistical_1999}.} 
Another important aspect, often neglected, is that different topological metrics may exhibit a non-trivial mechanistic correlation related to the way they are mathematically defined~\rev{\cite{Costa2007Characterizationcomplexnetworks,Lynall2010Functionalconnectivityand,Joyce2010newmeasurecentrality,Ekman2012PredictingErrors}}. Thus, when performing inference on several graph indices, their correlation should be accounted for. If multivariate statistical models are used, one solution is to assess the degree of multicollinearity, for example by computing variance inflation factors, and if necessary to adopt a shrinkage approach (e.g. ridge regression), or a modern regularized regression
framework such as the elastic net~\cite{Zou2005RegularizationElasticNet}. 
As in other statistical designs, when a series of independent hypothesis tests is performed on the same brain graph index (e.g. local-efficiency across different frequency bands or conditions), then multiple testing procedures, like Bonferroni corrections, should be considered to adjust the statistical significance of the result. When these multiple tests are dependent (e.g. node degree across different brain regions) then less restrictive procedures can be used, such as Benjamini and Hochberg's false discovery rate (FDR)~\cite{Benjamini1995ControllingFalseDiscovery}.

Machine learning approaches, represent a potential solution in that they can deal with both correlated and non-Gaussian variables. Under this type of approach, each brain graph is said to be embedded in a vector space, where the vector components are various topological metrics. Then, cross-validation is used to estimate the generalization ability of the machine learning
classifier. This type of approach is seeing increasing use~\cite{Cecchi2009DiscriminativeNetworkModels,Richiardi2011ClassifyingConnectivityGraphs,Bassett2012Alteredrestingstate}, with some state-of-the-art results obtained with this technique~\cite{Ekman2012PredictingErrors}, and clinical prediction starting to
be quite successful~\cite{Castellanos2013Clinicalapplicationsfunctional}. Finally, it is worth noting that graph metrics are not complete invariant (i.e. non-equivalent networks could have the same topological property), so the use of several measures 
in a multivariate setting can be justified for brain graph analysis as a way to alleviate invariant degeneracy~\cite{Bonchev1981IsomerDiscriminationby}.
\section{HEALTHY AND PATHOLOGICAL BRAINS}
\rev{While functional brain connectivity is known to be altered by many diseases~\cite{fox_clinical_2010,Guye2010GraphtheoreticalAnalysis,
Zhou2012Predictingregionalneurodegeneration,
Castellanos2013Clinicalapplicationsfunctional}, its computation should be done with caution because} clinical patients can have several structural, functional, and behavioral
alterations that may confound the subsequent graph analysis. A same alteration
can have a more or less severe confounding effect depending on the neuroimaging modality.
Structural brain deformations, for example, can cause uncertainty in EEG source localization
using individual head models derived from structural MRI, although it has been reported
that the effect may be negligible \cite{vonEllenrieder2006Effectsgeometrichead}.

Several pathologies, such as tumors, multiple sclerosis, or traumatic brain injury,
are accompanied by lesions in a variety of brain regions. This can cause
problems for structural segmentation and normalization algorithms, since the size,
localization and precise shape of lesions can vary unpredictably. In turn, this impacts 
the definition of the brain nodes and their inter-subject correspondence. Several solutions are available,
including acquiring structural images with different contrasts and using a multimodal segmentation
algorithms~\cite{Menze2010generativemodelbrain}, or a disease-specific/\rev{disease-adapted} segmentation procedure\rev{\cite{Seghier2008Lesionidentificationusing,VanLeemput2001Automatedsegmentationmultiple}}.
Localizing lesions, one can compare normalization results with and without lesions masked out,
and make a decision based on the observed added empirical error. Nevertheless, the assumption
that nodes in one brain graph match the nodes in another brain graph might
be misleading, and group or inter-individual comparisons should be approached with care.

Localized gray matter atrophy, or partial volume effects, may also result
in the segmentation algorithm assigning fewer voxels to a particular gray matter
region. For atlas-based assignment of cerebral regions to brain graph nodes, this may result in fewer voxels being assigned to a node and lead to an overall apparent decrease of FC between that node and the rest of the brain graph~\cite{Salvador2008simpleviewof}. This is due to the averaging operation commonly
used to summarize multiple signals within a macro area~\cite{Achard2011fMRIFunctionalConnectivity}.
Here, one solution is to regress out the gray matter density from each region prior to analysis, or to use robust FC methods~\cite{Achard2011fMRIFunctionalConnectivity}. More generalized atrophy, possibly concomitant with ventricle enlargement, or other 
large-scale deformations that cannot be modeled by affine transformations, can also induce normalization
and segmentation problems. A possible way to avoid this problem, is to use a normalization method with sufficient degrees
of freedom~\cite{Ashburner2007fastdiffeomorphicDARTEL}, or a non-linear method, to
warp the data into a common space. This approach can be successful even on severely deformed
brains~\cite{Achard2012Hubsbrainfunctional}. Similarly, if an atlas is used to define the brain graph nodes,
but the original population used to define the atlas does not correspond to the population under study (a typical case being elderly populations, since reference atlases are normally defined on young adults), these methods can be used to learn
a group-specific template from the available data.

Several brain pathologies have a vascular component, which adds to the natural spatial variability 
of the hemodynamic activity. For example, Alzheimer-type dementia is known to alter dilation amplitude
of blood vessels, and to some extent the dynamics of dilation in a regionally-specific fashion~\cite{Cantin2011Impairedcerebralvasoreactivity}. This is likely to disrupt FC measures
based on the BOLD signal. Using a deconvolution approach, with a flexible hemodynamic model, may be a way to improve the signal-to-noise ratio, although this type of approach is relatively new~\cite{Karahanoglu2013Totalactivation:fMRI}.

Finally, confounds can also come from group differences in motion (both for fMRI and EEG modalities), stress, or other physiological parameters such as cardiac or respiratory activity in fMRI imaging. However, a number of preprocessing techniques are available to solve these problems\rev{~\cite{Murphy2013RestingstatefMRI}}, and new alternative data-driven approaches are emerging~\cite{Marx2013novelapproachglobal,Churchill2013PHYCAA+}.
%
\section{ABSTRACTION LEVELS}
\label{s:abstract}

Much of our understanding of the brain functioning rests on the way it is measured and modeled~\cite{park_structural_2013}. In this sense, the ability to describe multivariate neural processes by looking at their network topology makes graph analysis unique compared to previous univariate methods that simply looks at the activity in single parts of the brain (e.g. power), or bivariate methods looking at pairwise FC (e.g. cross-correlation). One aspect, often ignored, is that there is a relation between univariate, bivariate and multivariate methods. Specifically, graph analysis depends on FC (i.e. graph theory applies to connectivity patterns), which in turn depends on activity (i.e. functional connectivity applies to brain signals).  

From a conceptual point of view, these different methods can be regarded as increasing abstraction levels of the original neural process (Fig. 2). As the abstraction level increases, new complementary information is obtained, but, at the same time, we bear off the intuitive interpretation of the results. When a change occurs in the original neural mechanism it propagates through different levels (forward analysis). Due to the increasing abstraction complexity the final changes in the graph metrics can be associated with changes in the FC measures but not directly to changes in the brain signals (backward interpretation). For example, suppose we are studying neural correlates of epileptic seizures using EEG signals (i.e. univariate), Granger-causality (i.e. bivariate), and the small-world graph index (i.e. multivariate). Then, it will be fundamentally incorrect to state that brain graphs increase their small-worldness during seizures. Despite its simplicity and impact, such a sentence skips two abstraction levels (i.e. bivariate and univariate) and can generate confusion and/or misinterpretation. Instead, a more appropriate statement should report an increase of small-worldness in the information propagation between the neuroelectrical activities during the seizure. This does not mean that summary interpretation of brain graph indices is always wrong, but it should be carefully used and possibly accompanied by complete explanation. Graph analysis of functional brain connectivity represents the last fundamental block of a processing pipeline that, if conceived and applied properly, can truly improve our comprehension of the brain functioning.

\section{FUTURE CHALLENGES}
Despite the promising advances of brain graph analysis, many methodological issues still remain unaddressed (see previous sections), thus limiting the general applicability of graph modeling and interpretation of its outcomes. In the following, we have identified three major issues that, in our opinion, represent the most pressing future challenges.
\subsection{Brain graph filtering}
As mentioned in Section~\ref{s:threshold}, graph filtering is one of the most crucial and recurrent issues for the thresolding of functional brain networks~\cite{Rubinov2010Complexnetworkmeasures}. The absence of an objective criterion to select a threshold $T$ for filtering the connectivity matrix often obligates researchers to repeat their analysis over a range of several increasing thresholds. This highly time-consuming procedure represents a critical problem when the size of the brain graphs, the number of subjects and/or conditions increases due to experimental requirements~\cite{langer_problem_2013}. Furthermore, this heuristic practice is not theoretically grounded and there is no hypothesis on the existence of a subinterval of thresholds for which the computed graph metrics remain relatively stable. 
However, finding a general criterion for selecting a possibly optimal threshold is not as trivial. There are many parameters to take into account like the graph size (number of nodes) and its topology (from lattice to random)~\cite{van_wijk_comparing_2010}. Other factors, like the FC method, the neuroimaging technique, and the experimental condition, could also influence the identification of an objective threshold. 
So far, only very few works have tried to address this issue based on minimum spanning tree approaches~\cite{AlexanderBloch2010DisruptedModularity} or data mining techniques~\cite{zanin_optimizing_2012}. However, these approaches present some drawbacks, i.e. minimum spanning trees are not unique while the proposed data mining-based technique still needs to explore a range of thresholds.
Certainly, a good threshold should neither filter all the possible links nor let them all survive. Indeed, analyzing brain graphs that are either almost empty or fully connected appears to be worthless~\cite{Achard2007Efficiencyandcost}.

\subsection{Statistical variability of brain graphs}
As mentioned in~\ref{s:abstract} brain graph indices are function of FC measures, which in turn are function of the measured (noisy) brain activity (Fig. 2). This introduces an important issue that has been completely neglected concerning the evaluation of possible error propagation across the forward analysis path. 
Currently, there is no predefined way to assess the statistical variability of topological metrics, which are extracted from connectivity patterns with unknown probability distributions. Although it appears crucial for outcome interpretation, only few works have partially addressed this statistical issue from a statistical network point of view~\cite{tremblay_bootstrapping_2013,fallani_nonparametric_2014}, while a complete framework investigating estimation errors from the measured brain activity to the graph indices is still lacking. However, providing confidence intervals will be fundamental in assessing the significance of the obtained results and proving the reliability of graph analysis of functional brain networks, especially in the clinical neuroscience. In this regard, possible validation frameworks can also come from the use of in-vivo optogenetic techniques for calcium imaging~\cite{wyart_let_2011}. In this case, the level of insertion loss is significantly attenuated providing signals with high signal-to-noise ratio and single-neuron resolution that can be considered as uncontaminated activity and exploited to validate error propagation across the forward analysis path (Fig. 2).

\subsection{Spatio-temporal brain graphs}
A last aspect concerns the evaluation of how dynamic changes in network connectivity relate to brain (dys)function. Functional brain networks are dynamic and change across time, on both long-time scales (eg, cortical plasticity after braid damages~\cite{Wang2010Dynamicfunctionalreorganization}) as well as short-time scales (eg, cognitive/motor learning~\cite{bassett_dynamic_2011}). 

It turns out that the development of methods to characterize dynamic networks is becoming critical in the area of data representation and analysis of functional brain connectivity. Previous studies have put some efforts in approaching the problem by simply tracking the temporal profile of individual topological metrics~\cite{fallani_cortical_2008, valencia_dynamic_2008, Wang2010Dynamicfunctionalreorganization}. While these studies have the merit to address a central issue, they still represent an over-simplified methodological approach and leave space to advances. In particular, possible solutions can come from the development of a new method that conceptualize time-varying brain graphs as multilayer networks. According to this generalized framework each layer models the interactions of the system at the time $t$, and the extracted time-varying graph metrics quantify the evolution of the topological properties across time~\cite{mucha_community_2010, tang_small-world_2010}.

Furthermore, the availability of many neuroimaging modalities with different spatial resolution, from voxel (fMRI) to sensor (EEG/MEG) levels, opens interesting opportunities to integrate the FC between brain regions across multiple spatio-temporal spatial scales. Limited attempts to include, separately, the presence spatial and time constraints, into a network description have already appeared in the literature, in various contexts~\cite{barthelemy_spatial_2011,holme_temporal_2012}. 
However, a comprehensive theory of dynamic networks with multiple node and link features is still lacking. A future challenge is then to provide a coherent theoretical framework to study and model multi-level and multi-dimensional brain graphs in terms of multi-layer networks embedded in space and time. 

\section{THE ROLE OF TECHNOLOGY}
At outlined in the present review, graph analysis of functional brain networks remains a complex tool that requires some methodological expertise. This aspect limits the impact on clinical neuroscience, where the need for simple and fast analytic tools is paramount to gain adoption in clinical routine. In this sense, recent EU-funded projects have attempted to bring brain image and signal processing tools on a distributed GRID environment which is accessible from different hospitals and certified users across the world\footnote{\url{https://www.eu-decide.eu/}}$^,$\footnote{\url{https://neugrid4you.eu/}}. These efforts, aiming to promote the use of analytic tools among the clinical neuroscience community, will be fundamental in receiving feedback and improving the applicability of graph analysis of functional brain networks.

In the meantime, new results are accumulating rapidly and we are still facing the issue of how to integrate this huge amount of findings to achieve a common consensus on the specific functional network properties concerned by different brain diseases. 
This is in part due to the existence of many methods and parameters that can be used and tuned to measure and characterize brain graphs (Fig. 1). This high number of degrees of freedom in the data analysis inevitably increases the dispersion of the results and affects their convergence. In this regard, the role of comprehensive meta-analysis methods will be fundamental in the near future to create a ``dictionary'' associating different brain diseases with the most relevant brain graph topological changes. Notably, some efforts have been recently done in this direction resulting in a very useful and comprehensive review for stroke disease~\cite{grefkes_reorganization_2011}. 

From a broader perspective, international initiatives, like the European Human Brain Project (HBP)\footnote{\url{https://www.humanbrainproject.eu/}}, the American Brain Research through Advancing Innovative Neurotechnologies (BRAIN) initiative\footnote{\url{http://www.nih.gov/science/brain/index.htm}}, or the Chinese Brainnetome project\footnote{\url{https://www.brainnetome.org/}} aiming to combine new knowledge from different neuroimaging centers and provide access to data sources, platforms, and infrastructures, can be fundamental in accelerating such progress. Furthermore, these projects will address an ambitious challenge that is giving a complete mapping of brain activity. Indeed, one of the most limiting factor that impedes the description of all the mechanisms underlying physiological and/or pathological behavior is that we do not yet possess the technological means to access the activity of each single element (i.e. neuron) of the system~\cite{alivisatos_brain_2012}. Independently on the success of these projects, the development of network-based methods to analyze the interaction between distributed neuronal signals recorded at micro, meso or macro levels is paramount.~\rev{These analytic tools can be useful to quantify the topology of a hugely complex brain networks emerging from the connectivity of more than 80 billion neurons~\cite{azevedo_equal_2009}. Moreover, they can provide abundant models to describe the basic interaction between specific neuronal ensembles and predict network changes correlated to cognitive/motor behavior and disease.
}

\section{CONCLUSIONS} 
In the last decade, we have witnessed an impressively increasing number of neuroscience studies using graph analysis of functional brain networks. Notably, clinical neuroscientists have hypothesized for years that behavioral cognitive/motor impairment related to neurological diseases was multifaceted and could be caused by a dysfunction involving many interacting remote regions rather than a single area. In this sense, the recent arrival of analytic tools capable of modeling and characterizing brain activity from a system perspective has represented a unique opportunity towards a more complete understanding of brain diseases and towards the development of effective network-based neuromarkers for early diagnosis and prognosis.  

This evidence explains why the field of network science has gained greater popularity in the clinical neuroscience and why, at the same time, the risk of a rush towards its frenetic and counterproductive application becomes more and more concrete. The major thrust of the present review is to provide researchers and neuroscientists with focused indications to make sense of their functional brain network analysis and avoid the most common traps in using graph theoretical approaches.

\section*{Acknowledgments}
F. De Vico Fallani is financially supported by the French program ``Investissements d'avenir" ANR-10-IAIHU-06. 
J. Richiardi is supported in part by a Marie Curie International Outgoing Fellowship of the European Community's 7th framework programme (PIOF-GA-2011-299500). M Chavez is partially supported by the EU-LASAGNE Project, Contract No.318132 (STREP). 

\bibliography{manuscript}
\bibliographystyle{royalb}

\begin{table}[htbp]
\centering
\begin{tabular}{m{1cm} l c c c c} \toprule
& Method & \multicolumn{3}{c}{Features} & Requirements \\ 
\cmidrule(rl){3-5} \cmidrule(rl){6-6} 
&  & Nonlinearity & MVAR & Frequency & Parametric\\
\midrule
\multirow{3}{*}{\rotatebox[origin=c]{90}{Undir.}} & Correlation & no & partial correlation~\cite{marrelec_partial_2006} & spectral coherence~\cite{carter_coherence_1987} & no\\
& Mutual information & yes & yes & no &  yes\\
& Phase coherence & yes & no & yes & no\\
\midrule
\multirow{3}{*}{\rotatebox[origin=c]{90}{Dir.}} & Lagged correlation & no & no & phase difference~\cite{rosenblum_detecting_2001} & time lag\\
& Transfer Entropy & yes & yes & no & time lag\\
& Granger-causality & nonlinear GC~\cite{marinazzo_nonlinear_2011} & partial GC~\cite{guo_partial_2008} & PDC~\cite{baccala_partial_2001}, DTF~\cite{kaminski_new_1991} & model order\\
\bottomrule
\end{tabular}
\caption{List of the basic principles implemented in the most used functional brain connectivity (FC) methods. Methods are organized according to their ability to characterize undirected or directed connectivity. These methods were initially developed to detect linear, bivariate, and time-domain interactions between brain signals. Extensions to nonlinear, multivariate, frequency domain are indicated in the Features section, together with their main methodological requirements. A complete review of these methods can be found in~\cite{sakkalis_review_2011} (EEG/MEG) and~\cite{rogers_assessing_2007} (fMRI).}
\label{t:connMeasures}
\end{table}

\begin{table}[htbp]
\centering
\begin{tabular}{m{1cm} l c c  c c c c} \toprule

& & \multicolumn{5}{c}{Nb Samples}\\
 & & 100  & 200 & 300 &  400 & 500 \\
\cmidrule(rl){3-7} 
\multirow{4}{*}{\rotatebox[origin=c]{90}{Nb Nodes}} & 50 & 0.865 & 0.686 & 0.589  & 0.519 & 0.472 \\
& 100 & 0.89 & 0.721 & 0.624  & 0.552 & 0.504 \\
& 500 & 0.928  & 0.783 & 0.689  & 0.617 & 0.567 \\
& 1000 & 0.939 & 0.804 & 0.712  & 0.64 & 0.589 \\
\bottomrule
\end{tabular}
\caption{Wavelet correlation values for scale 3 significantly different from 0 when applying FWER control (for a sample frequency equal to 1.1~Hz, this corresponds to Pearson correlation computed at frequency range equal to $[0.06-0.11]$~Hz \cite{Achard2006JNeuro}). Each value in the table correspond to the smallest correlation that can be considered to be significantly different from zero given a number of brain nodes and a number of time points for the respective signals. For example, with 500 nodes and 500 time points, correlations greater than 0.567 are considered to be significantly different from 0 at wavelet scale 3.}
\label{t:filterGraph}
\end{table}
\efloatseparator

\begin{table}[htbp]
\centering
\begin{tabular}{m{1cm} l c c c c c} \toprule
& Metric &  \multicolumn{3}{c}{Features} & Requirements \\
\cmidrule(rl){3-5} \cmidrule(rl){6-6}
&  & Weighted & Directed & Negative & Connected\\

\midrule
\multirow{6}{*}{\rotatebox[origin=c]{90}{Large}}


& characteristic path length & yes & yes  & no & yes \\

& global-efficiency & yes & yes  & no & no \\

& clustering coefficient & yes & yes  & no  & no \\

& local-efficiency & yes & yes  & no & no \\

& modularity & yes & yes  & yes & no \\



\midrule
\multirow{5}{*}{\rotatebox[origin=c]{90}{Intermed.}}

& communities & yes & yes  & no &  no \\

& motifs & yes & yes  & no &  no \\

& edge betwenness & yes & yes  & no &  no \\

& redundancy~\cite{de_vico_fallani_redundancy_2012} & no & yes  & no & no \\
 
\midrule
\multirow{5}{*}{\rotatebox[origin=c]{90}{Small}}

& degree & yes & yes  & yes & no \\

& node betweenness & yes & yes  & no & no \\

& eigenvector centrality & yes & yes  & yes & yes \\ 

& accessibility~\cite{chavez_node_2013} & yes & yes  & no & yes \\

\bottomrule
\end{tabular}
\caption{List of methods implemented in the most used and recent brain graph metrics (or indices). Methods are organized according to their ability to characterize large (whole network), intermediate (sub-networks), or small (nodes) topological scales. These methods were initially developed to describe the topology of networks with unweighted (binary), undirected, and positive weight links. Extensions to weighted, directed, negative weights are specified in the Features section, together with their main methodological requirements. An exhaustive review of these methods can be found in~\cite{Rubinov2010Complexnetworkmeasures}.}
\label{t:graphMetrics}
\end{table}

\begin{figure}
\centering
\includegraphics[width=1\textwidth]{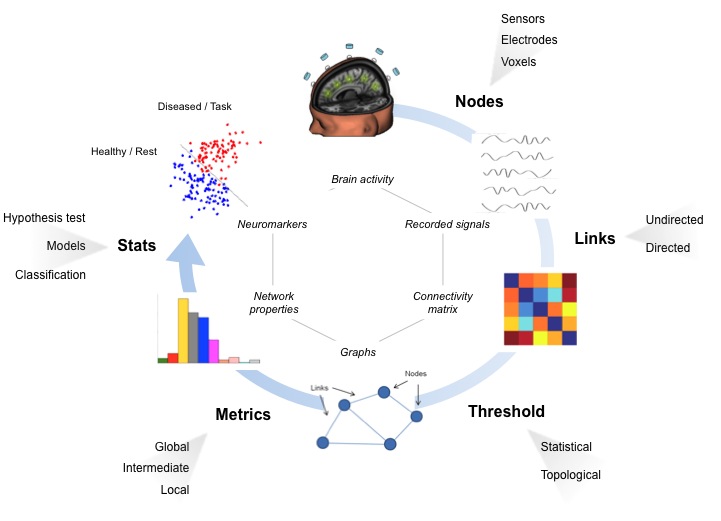}
\caption{Processing pipeline for functional brain connectivity modeling and analysis. Nodes correspond to specific brain sites according to the used neuroimaging technique -> \textit{Section~\ref{s:nodes}}. Links are estimated by measuring the functional connectivity (FC) between the activity of brain nodes; this information is contained in a connectivity matrix -> \textit{Section~\ref{s:links}}. By means of filtering procedures, based on thresholds, only the most important links constitute the brain graph -> \textit{Section~\ref{s:threshold}}. The topology of the brain graph is quantified by different graph metrics (or indices) that can be represented as numbers (e.g. the colored bars) -> \textit{Section~\ref{s:index}}. These graph indices can be input to statistical analysis to look for significant differences between populations/conditions (e.g. red points correspond to brain graph indices of diseased patients or tasks, blue points stand for healthy subjects or resting states -> \textit{Section~\ref{s:stats}}.}
\label{Figure 1}
\end{figure} 

\begin{figure}
\centering
\includegraphics[width=0.55\textwidth]{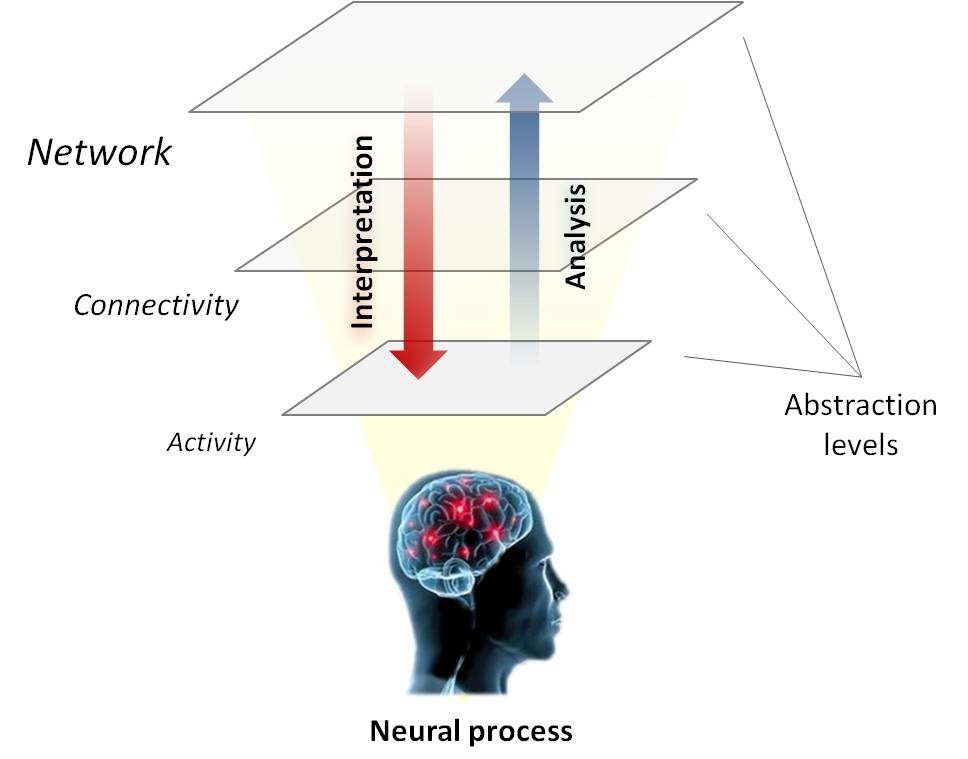}
\caption{Abstraction levels for the analysis and interpretation of brain graphs. \textit{Forward analysis} (blue arrow): changes in the neural process generates modifications in the measured brain activity (unvariate analysis -> abstraction level 1). Functional connectivity (bivariate analysis -> abstraction level 2) applies to the measured brain signals, and graph modeling (multivariate analysis -> abstraction level 3) applies to functional connectivity. \textit{Backward interpretation} (red arrow): results obtained with graph analysis (abstraction level 3) refer to the previous abstraction level, i.e. functional connectivity, and can be directly associated neither to changes measured in the brain activity (abstraction level 1) nor to the original neural process. Accordingly, the interpretation of the results obtained with graph analysis is mediated by the choice of the functional connectivity measure and by the used neuroimaging technique.}
\label{Figure 2}
\end{figure} 

\end{document}